\documentclass[aps,twocolumn,pra,showpacs]{revtex4}
\usepackage{amssymb}
\usepackage{mathrsfs}
\usepackage{graphicx}
\usepackage{amsmath,amssymb,amsthm,bbm,latexsym}
\usepackage{amsthm}
\usepackage{amsbsy}
\usepackage{epsfig}
\usepackage{graphicx}
\usepackage{float}
\usepackage{dcolumn}
\usepackage{amsmath}

\begin{document}

\author{Zhao-Ming Wang$^{1,2}$\thanks{%
Email address: mingmoon78@126.com}, Lian-Ao Wu$^{2}$\thanks{%
Email address: lianao_wu@ehu.es}, Jun Jing$^{3}$\thanks{%
Email address: Jun.Jing@stevens.edu}, Bin Shao$^{4}$, Ting Yu$^{3}$\thanks{%
Email address: Ting.Yu@stevens.edu}} \affiliation{$^{1}$ Department
of Physics, Ocean University of China, Qingdao, 266100, China}
\affiliation{$^{2}$ Department of Theoretical Physics and History of
Science, The Basque Country University(EHU/UPV) and IKERBASQUE,
Basque Foundation for Science, 48011 Bilbao, Spain}
\affiliation{$^{3}$Center for Controlled Quantum Systems and
Department of Physics and Engineering Physics, Stevens Institute of
Technology, Hoboken, New Jersey 07030, USA} \affiliation{$^{4}$Key
Laboratory of Cluster Science of Ministry of Education, and School
of Physics, Beijing Institute of Technology, Beijing 10081, China}

\title{Non-perturbative Dynamical Decoupling Control: A Spin Chain Model}

\begin{abstract}
This paper considers a spin chain model by numerically solving the exact model to explore the
non-perturbative dynamical decoupling regime, where an important issue arises recently \cite{Jing2012}.
Our study has revealed a few universal features of non-perturbative dynamical control irrespective of the
types of environments and system-environment couplings. We have shown that, for
the spin chain model, there is a threshold and a large pulse parameter region
where the effective dynamical control can be implemented, in contrast to the
perturbative decoupling schemes where the permissible parameters are
represented by a point or converge to a very small subset in the large
parameter region admitted by our non-perturbative approach. An important
implication of the non-perturbative approach is its flexibility in implementing
the dynamical control scheme in a experimental setup. Our findings have
exhibited several interesting features of the non-perturbative regimes such as
the chain-size independence, pulse strength upper-bound, noncontinuous valid
parameter regions, etc. Furthermore, we find that our non-perturbative scheme
is robust against randomness in model fabrication and time-dependent random
noise.
\end{abstract}

\maketitle

\section{Introduction}
Central to quantum science and technology is to combat decoherence caused by
inevitable external noises or quantum operation inaccuracies \cite{zurek2003}.
Strategies for controlling a quantum state include the quantum error correction
codes \cite{shor1995,ekert1996,gottesman1996} and dynamical coupling control
such as fast-strong pulses (``bang-bang'') control
\cite{Knill2000,Lidar1998,Viola2,Uhrig1,Uhrig2,Lidar,Wu09}. One common feature
of these dynamical control approaches is that almost all the theories are
defined in a perturbative manner. Despite the progress made in theoretical
studies and experimental realizations, the perturbation theories still lack a
satisfactory understanding of their validity and applicability, especially
the divergence issue due to the onset of a possible phase transition. Recently,
a careful examination of the dynamical decoupling or bang-bang control has led
to new breakthroughs achieved through introducing a non-perturbative dynamical
decoupling scheme based on exact stochastic master equations in bosonic baths
\cite{Jing2012}. By using this scheme, we showed that there is a threshold and
a large pulse parameter region where the effective dynamical control can be
implemented, in contrast to the perturbative decoupling schemes where the
permissible parameters converge to a point or form a very small subset in the
large parameter region admitted by our non-perturbative approach. Our
non-perturbative approach has shown impressive flexibility in implementing the
dynamical control scheme in experimentation.

This paper uses a spin chain model to explore the non-perturbative dynamical
decoupling regime. Our results demonstrate, through new type of environment and
new numerical methods, several universal features of the non-perturbative
dynamical decoupling emerge. More specifically, we show that there is a
threshold and a large pulse parameter region where the effective dynamical
control can be implemented.

We model the first spin in an $XY$-type model as our system of interest and the
rest as our environment. The bath spectral density is chosen as a complicated
function indicating that the ratios between two arbitrary levels are normally
not rational numbers. We have derived an exact ``master equation'' for our spin
system. It should be noted that for the spin environment, the environmental
correlation function is very different from that for a bosonic model. However,
our results obtained from this particular $XY$-type model are expected to be
applicable to other spin-environment models. Our numerical calculations show
that effective control can be made in many different ways either through a
multiple pulse sequence, as in the case of bang-bang control, or just a single
pulse applied within the bath memory time. The effectiveness of the dynamical
control is not sensitive to the size of spin chain, an observation that
showcases the universality of our findings. For the spin chain model, we have
shown that there is an upper-bound for pulse strength in non-perturbative
regimes when the pulse duration is sufficiently large, which is rather
counter-intuitive from the point of view of the strong fast pulse control
(bang-bang). We also consider the time-independent randomness for coupling
constants and site energies of the $XY$-type spin chain. We show that our
results and conclusion are robust to the random couplings and site energies
even in the strong-coupling limit. Finally, we consider the time-dependent
random noise, which may simulate the noisy effects due to additional
environmental variables. We show that the quality of dynamical control in the
non-perturbative regime is also robust to the influence of the time-dependent
noise.

\section{The model}

The spin chain model considered here is represented by the following
Hamiltonian,
\begin{equation}
H=H_{S}+H_{B}+H_{SB}
\end{equation}
where $H_{S}$ and $H_{B}$ is the system's and environment's Hamiltonian.
$H_{SB}$ is the interaction between the system and environment. We consider an
$XY$-type spin-chain model, where the first spin of the chain is our system of
interest and the environment is composed of all the other $N-1$ spins. The
system Hamiltonian and control are $H_{S}=c(t)Z_{1}$, where $c(t)$ is a
time-dependent control function. The control is only applied to the system, and
the $c(t)$ is chosen as a sequence of periodic rectangular pulses:
\begin{equation}
c(t)=\bigg\{%
\begin{array}{c}
\Psi \text{ \ \ }n\tau  <t<n\tau +\Delta,n\text{ is integer} \\
0,\text{ \ \ \ \ \ \ \ \ \ \ \ \ \ \ \ otherwise\ \ \ \ \ \ \ \ \ \ \ \ }%
\end{array}%
\end{equation}
where $\Psi$ is the pulse strength, $1/\tau$ is the pulse frequency and
$\Delta$ is the width of the pulse in one period. The three parameters span the
parameter space of the non-perturbative dynamical decoupling scheme as in
\cite{Jing2012}. The limiting case with $1/\tau, \Delta\rightarrow 0$ and $\Psi
\rightarrow \infty$ corresponds to the idealized Bang-Bang control.

The $XY$-type spin-chain Hamiltonian is modeled as
\begin{eqnarray}
H_{SB}&=&-J_{1,2}(X_{1}X_{2}+Y_{1}Y_{2}), \\
H_{B}&=&-\sum_{i=2}^{N-1}J_{i,i+1}(X_{i}X_{i+1}+Y_{i}Y_{i+1})+H_{\text{site}}
\end{eqnarray}
where $J_{i,i+1}$ is the exchange interaction between sites $i$ and
$i+1$. The system and environment coupling constant is $J_{1,2}$.
$X_{i},Y_{i}$ denote the Pauli operators acting on spin $i$, and $N$
is the total number of sites. We consider a natural configuration of
the spin chain with open ends. The $z$-component of the total spin
is a conserved quantity implying the conservation of the total
excitations in the system. For simplicity, we consider that the
system is in the ``one-magnon'' state, in which the total number of
up spins is one and $J=J_{i,i+1}=1.0$. $H_{\text{site}}$ is the site
energy and will be specified in our later discussion. Our numerical
calculations show that, within the ``one-magnon'' subspace, the sign
of $J$ (plus or minus) does not affect the results.

The system dynamics alternates between control and free evolution. We start
with a control pulse with the width $\Delta$, followed by $\tau-\Delta$ free
evolution. The evolution operator of the whole system for the first time
interval or period will be $U_{0}(\tau -\Delta)U(\Delta)$, where $U(\Delta
)=\exp [-i\Delta H]$ and $U_{0}(\tau -\Delta)=\exp [-i(\tau -\Delta )H_{0}]$.
Here $H_0$ denotes the free evolution without control. We then repeat the same
operations from time $\tau$ to $2\tau$, $\cdots$. The bang-bang control theory
is the idealized limit, $\Delta ,\tau \rightarrow 0$ and $
\Psi \rightarrow \infty$, which has been shown to be able to eliminate the
interactions between the system and bath.

We can diagonalize the Hamiltonian $H$ and $H_{0}$ such that $H_{d}=W^{\dag
}HW,H_{0d}=V^{\dag }H_0V$. The evolution operators can therefore be expressed
by
\begin{eqnarray}
U(\Delta ) &=&W\exp [-i\Delta H_{d}]W^{\dag } \\
U_{0}(\tau -\Delta ) &=&V\exp [-i(\tau -\Delta )H_{0d}]V^{\dag }
\end{eqnarray}
Suppose that our system, the first spin, is initially in the state $\left\vert
\phi(0)\right\rangle =\left\vert 1\right\rangle $, while the other spins are in
the spin down state $\left\vert 0\right\rangle$. The spin chain will be in a
product state $\left\vert \Phi (0)\right\rangle =\left\vert 1\rangle \otimes
|00\cdots0\right\rangle= \left\vert 100\cdots0\right\rangle $.

After $m$-time control, the fidelity at the time $m\tau $, measuring the
survival probability of the initial state $\left\vert \phi (0)\right\rangle $,
can be defined as $F=\sqrt{\left\langle \phi (0)\right\vert \rho
(t)\left\vert \phi (0)\right\rangle }$, where $\rho(t)$ is the reduced density
matrix of the state at the first sites at $t=m\tau $. The $N$ eigenvectors and
eigenvalues of the Hamiltonian $H,H_{0}$ can be obtained by numerical
diagonalization.

In the next section, we will analyzes the quality of dynamical control in terms
of the fidelity defined above.

\section{RESULTS and DISCUSSIONS}

\begin{figure}[htbp]
\centering
\includegraphics[scale=0.8,angle=0]{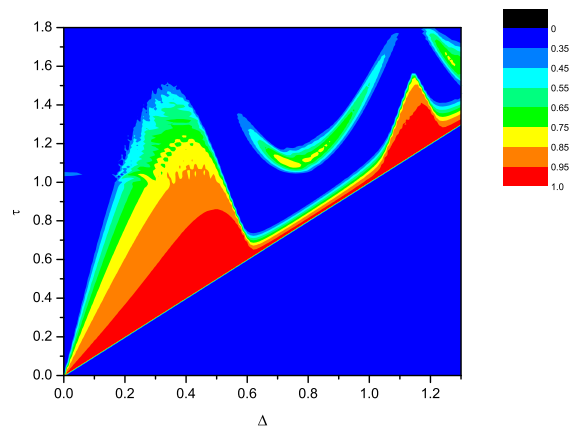}
\caption{(Color on line) The fidelity after 128-time control pulse
($t=128\tau$), the initial state is $\left\vert 1\right\rangle $ at the first
spin. The bottom triangular area has no physical meaning since $\tau$ is always
larger than $\Delta$ by definition. $\Psi=8.0$, $\emph{N}=130$.}
\label{fig:1}
\end{figure}

Suppose that our system, the first spin, is initially in the state $\left\vert
1\right\rangle$. With time evolution, the state will be distributed to the
other spins due to the interaction among the spin chain. The periodical
rectangle pulses will suppress the disturbing process, a well-known fact in the
case of bang-bang control. Fig. \ref{fig:1} plots the contour fidelity as a
function of the pulse period $\tau $ and pulse duration $\Delta $ at time
$t=128\tau $ (the $\Delta -\tau$ phase diagram). It clearly shows that, similar
to the dissipative bosonic bath \cite{Jing2012}, there is a large parameter
region where the dynamical decoupling (or noise suppression) works equally well
as shown in the red zone in Fig. \ref{fig:1}. The left-bottom corner($\Delta$
and $\tau \rightarrow 0$) represent the idealized pulses dynamical decoupling,
which has the same fidelity as the regimes with parameters $\tau$ and $\Delta$
in the red zone. Interestingly, there is an upper-bound for the parameter
$\Delta \approx 1.3$ in this spin model. The onset of the failure of an
effective control is the fundamental restriction of the bath memory time to be
discussed later.

\begin{figure}[htbp]
\centering
\includegraphics[scale=0.8,angle=0]{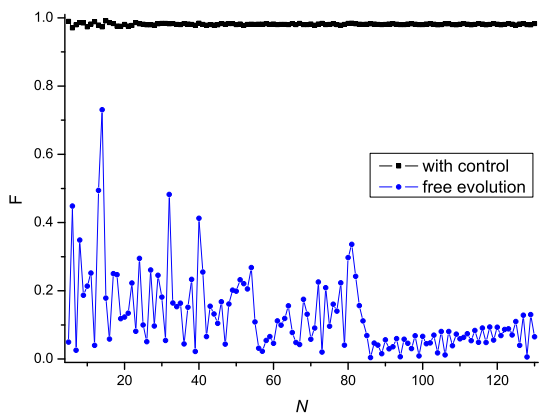}
\caption{(Color on line) The fidelity as a function of number of sites \emph{N}
under free and control Hamiltonian, where $\Psi=8.0,\Delta=1.2,\tau=1.3$ and
$t=128\tau$.} \label{fig:2}
\end{figure}

The analysis presented above is for the case when the number of sites,
\emph{N}=130. Since our environment contains a finite number of spins, it is
important to analyze the length dependence of our control scheme, in order to
eliminate the finite size effects. In Fig.~\ref{fig:2}, we plot the fidelity of
the system state at the time $t=128\tau$ for varied chain lengths or the
numbers of sites. We compare the fidelity of free evolution with the case that
the dynamical control is applied in this figure. The horizontal axes represents
the number of sites $N$ and the vertical is the fidelity. There are several
interesting new features in this figure. First, for the free evolution, while
fidelity is not good as expected, clearly the fidelity shows significant
oscillation for odd-even numbers of sites, whose magnitude decreases with $N$.
This is a typical finite-size effect. However, the quality of the
non-perturbative dynamical decoupling, in the non-perturbative control pulse
region with $\Delta =1.2,\tau =1.3$, does not depend on the chain length $N$,
as shown clearly in the figure. The fidelity is close to $0.98$ for all the
total numbers of spins $N$. This implies that the finite size effect of the
chain, dose not affect our results and conclusions.

\begin{figure}[htbp]
\centering
\includegraphics[scale=0.8,angle=0]{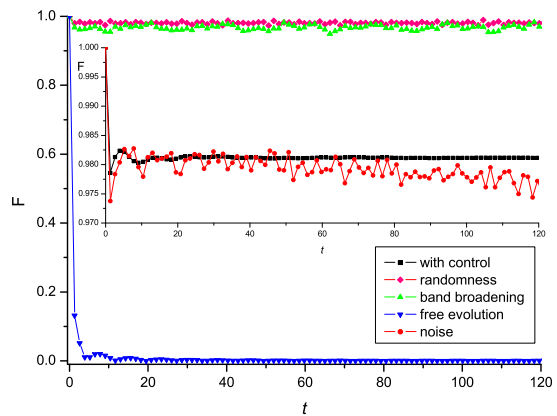}
\caption{(Color on line) The fidelity as a function of control times $t$.
$\Psi=8.0,\Delta=1.2, \tau=1.3, \emph{N}=130$. We explain these curves in the
text. }\label{fig:3}
\end{figure}

The above analysis is for the case with a fixed time $t=128\tau$.
Now we discuss the time evolution of the fidelity. Fig.~\ref{fig:3}
plots the fidelity versus time $t$. We compare the five cases: free
evolution, and controlled evolution under the one of the following
four conditions: (1) constant coupling $J$, (2) band broadening, (3)
randomness coupling $J+\gamma$rand$(i)$ and (4) stochastic coupling
$J+\eta$rand$(i)$. Here the band broadening refers to the
uncorrelated random potentials, which is defined by
$H_{\text{site}}=\epsilon\sum_{i}$rand$(i)$ ($\epsilon=0.5$ in the
figure) and generated in the process of physical fabrication.
$\gamma$rand$(i)$ is a random function of sites $i$, where $\gamma$
(taken as $0.5$ in the figure) is the magnitude of the random
function. rand$(i)$ denotes a stochastic number uniformly
distributed in the interval $(-1,1)$. It is noticeable that the
dynamical control quality, characterized by the fidelity, does not
change significantly with these two types of random distributions,
even in the long-time limit, which indicates that our
non-perturbative dynamical decoupling control is robust against the
defect or frustration in experimental fabrication of the spin
systems. Once the spin-chain is generated, it does not change with
time. However, it is important to consider a time-dependent random
noise due to an external agent (or environment). Equally
interesting, our numerical scheme allows us to model the noise as
the perturbation by $\eta$rand$(i)$ ($\eta=0.1$ in
Fig.~\ref{fig:3}), which changes randomly before each pulse period,
i.e. the random number $\eta$rand$(i)$ is constant within one period
$\tau$ but varies for different periods. Surprisingly, the control
quality is shown to be robust compared to the cases without the
time-dependent noises, as shown by the inset of Fig.~\ref{fig:3}.
Again, our analysis shows that another universal features of our
finding based on the non-perturbative dynamical decoupling scheme.

\begin{figure}[htbp]
\centering
\includegraphics[scale=0.7,angle=0]{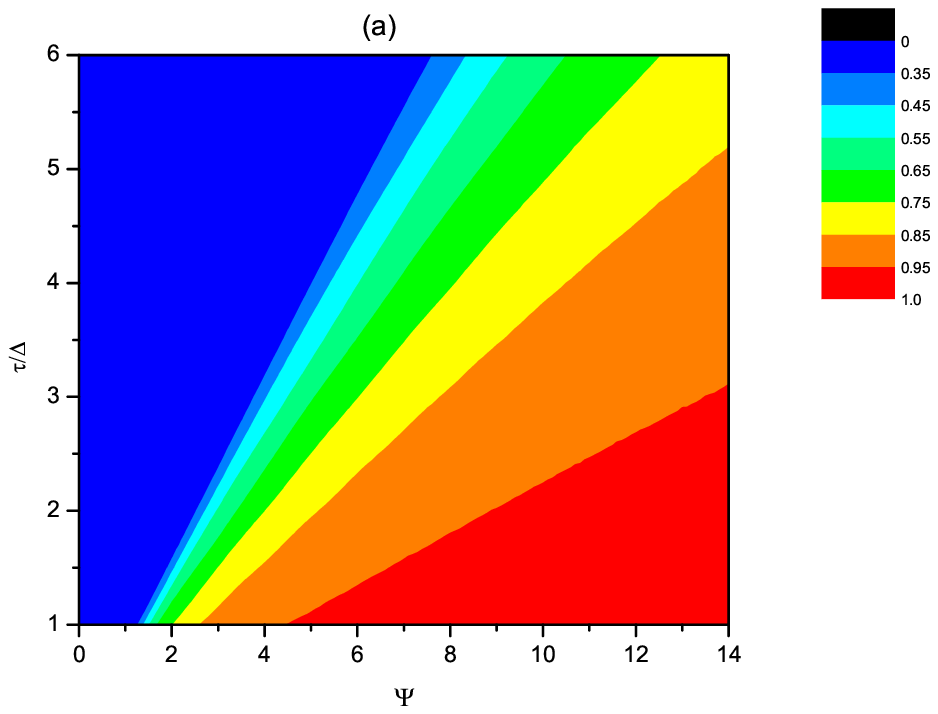}
\includegraphics[scale=0.7,angle=0]{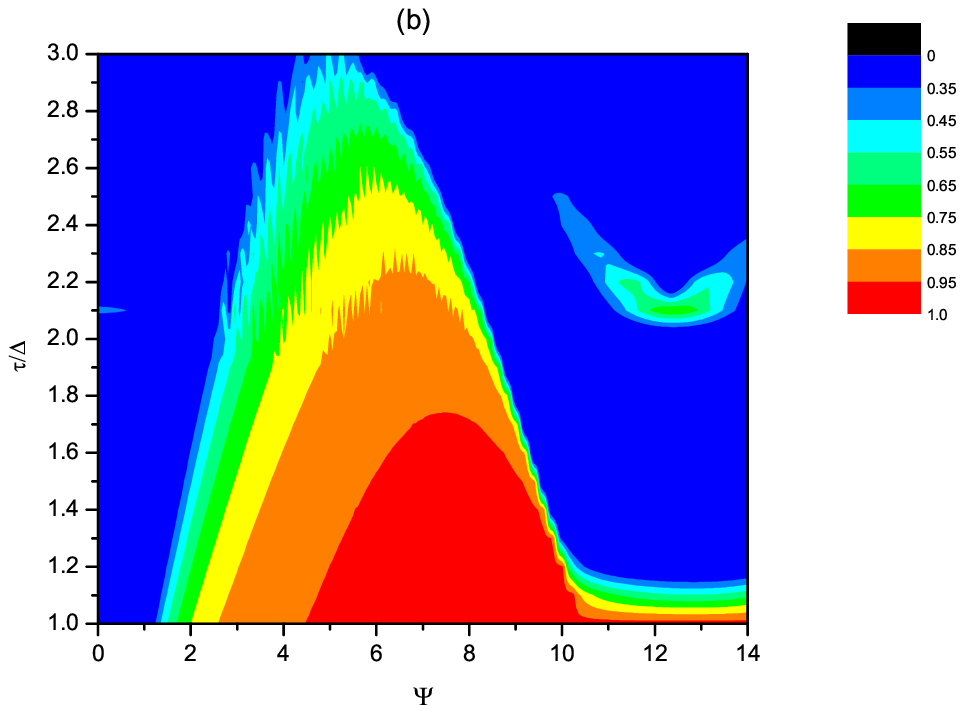}
\includegraphics[scale=0.7,angle=0]{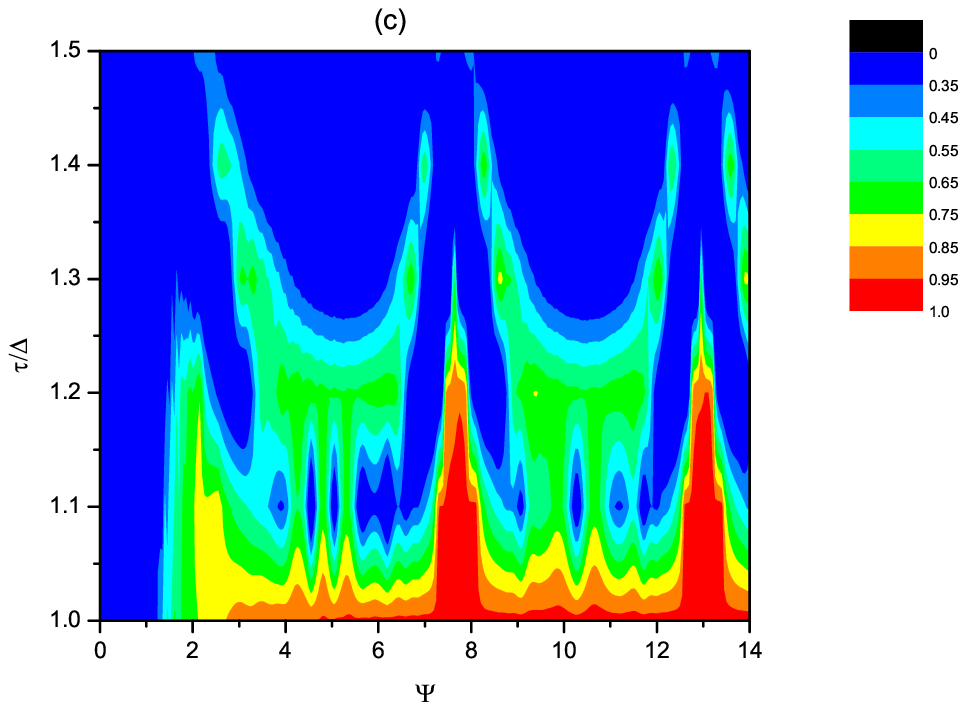}
\caption{(Color on line) $\tau/\Delta-\Psi$ phase diagram of the fidelity at a
control times m=128. $\emph{N}=130$. (a) $\Delta=0.1$, (b) $\Delta=0.5$, (c)
$\Delta=1.2$.}\label{fig:4}
\end{figure}

In Fig. \ref{fig:4}, we plot the contour fidelity as a function of pulse
strength $\Psi$ and $\tau /\Delta$ for given $\Delta =0.1,0.5,1.2$,
respectively. There is always a region where the dynamical control works with
high fidelity, as shown in the red zone where $F>0.95$. Surprisingly, there are
a lower-bound and an upper-bound for the strength in non-perturbative regimes.
For instance, when $\Delta =0.5$, this region is  $6.0<\Psi<9.0$ when $1.0<\tau
/\Delta <1.6$. When the parameter is close to the bang-bang regime, e.g.,
$\Delta =0.1$, there is only a lower-bound. In other words, only bang-bang
requires strong pulse strength, while non-perturbative control does not.

\section{Environment correlation by Feshbach projection-operator partitioning
technique}

Using $PQ$ partitioning technique \cite{Jing2012}, an $n$-dimensional wave
function $\psi_{t}$ can be divided into two parts: an interested
one-dimensional vector $P(t)$ and the rest $(n-1)$-dimensional vector $Q(t)$.
$\psi_{t},H_{eff}$ can be simply written as
\begin{equation}
\psi_{t}=\left[
\begin{array}{c}
P \\
Q%
\end{array}%
\right],\quad H_{eff}=\left[
\begin{array}{cc}
h & R \\
R^{T} & D%
\end{array}%
\right]
\end{equation}
where the $1\times 1$ matrix $h$ and $(n-1)\times (n-1)$ matrix $D$ are the
self-Hamiltonian living in the subspaces of $P$ and $Q$. For our model, $h$ is
the control function $c(t)$ and $R=\left[
\begin{array}{ccccc}
J & 0 & 0 & ... & 0%
\end{array}%
\right] $ is $1\times(n-1)$-dimensional.
\begin{equation}
D=\left[
\begin{array}{ccccc}
0 & J & .. & 0 & 0 \\
J & 0 & .. & 0 & 0 \\
.. & .. & .. & .. & .. \\
0 & 0 & .. & 0 & J \\
0 & 0 & .. & J & 0%
\end{array}%
\right]
\end{equation}
In the selected one dimensional subspace, $P$ satisfies
\begin{equation}\label{Pt}
i\overset{.}{P}=hP-i\int_{0}^{t}dsg(t,s)P(s)
\end{equation}
where in our case
$g(t,s)=g(t-s)=Re^{-iD(t-s)}R^{T}=RLe^{-id(t-s)}L^{\dag}R^{T}$, $d$ is
the diagonalized matrix, $d=L^{\dag }DL$. Specifically,
\begin{equation}
g(t-s)=J^2\sum\limits_{k}\left\vert L_{1k}\right\vert ^{2}e^{-iE_{k}(t-s)}
\end{equation}
where $L_{1k}$ are matrix elements of the first row and the $k$th column of the
matrix $L$ and $E_k$ is the $k$th eigenvalues of $D$.

\begin{figure}[htbp]
\centering
\includegraphics[scale=0.7,angle=0]{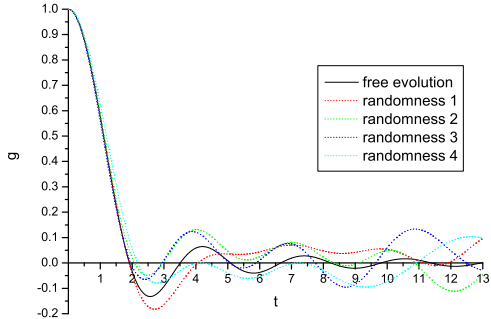}
\caption{(Color on line) The real part of bath correlation \emph{g} versus
\emph{t-s}, where $\emph{N}=130$ and $J=1$. The different dashed curves denote
different random distributions.}\label{fig:5}
\end{figure}

In general, Eq.~(\ref{Pt}) is a one-dimensional exact equation \cite{JWYY} for
arbitrary baths and has to be numerically solved. However, $g(t,s)$, is often
reduced to $g(t-s)$, is a bath correlation function and often has common
behaviors that hardly depend on the details of baths. Normally, the real part
of the function $g(t-s)$ starts from one and decays (rapidly) to zero with time
$t-s$, unless baths have limited numbers of frequencies, for instance, the
spectral distribution of a few Harmonic oscillators. The imaginary part
corresponds to an oscillating factor and is not responsible for the leakage
from the $P$ space, resulting in the renormalized energies. Fig.~\ref{fig:5}
shows the behaviors of the real part of the environment correlation function of
this specific model. The spectrum structure of $D$ in our model is sufficiently
complicated such as the ratios of any two energy levels are not rational
numbers. We see that the correlation function (memory function) vanishes at the
lifetime $\approx 1.7$, with a few small damping oscillations due to the finite
number of spins. Fig.~\ref{fig:5} also shows that the effect of randomness on
the correlation function ($\gamma$rand$(i)$) of $J$ (See the four dashed
lines). The randomness only affects the correlation function after lifetime and
different kinds of randomness yield the out-of phase curves, meaning that the
randomness may eliminate the possibility of the revival of the memory after the
lifetime. Strong and fast pulses suppress the decay rate and effectively
prolong the lifetime of the correlation function. This is one important
mechanism underlying the dynamical decoupling control (see, e. g., \cite{Wu09}
and references therein). On the contrary, our valid parameter region may allow
very wide and slow pulses, e.g. $\tau\approx 1.3$ (almost the lifetime), which
is obviously beyond the known perturbative suppression mechanism. Our
calculations confirm further that the valid parameter $\tau$ should be chosen
within the lifetime. That is, the dynamical control is effective when the
environment is still in non-Markovian regimes.

\section{CONCLUSIONS}

In conclusion, we have investigated a non-perturbative approach to dynamical
control theory through an $XY$-type spin model. We have shown that dynamical
control with high fidelity can be achieved in a large control parameter region.
There are many interesting features arising from this spin-chain model
including the chain-size independence, the pulse strength upper-bound,
noncontinuous valid parameter regions, etc.

Our initiative has opened the new avenue for exploring the dynamical decoupling
scheme in a more realistic situation allowing non-idealized pushes. Our results
suggest that the non-perturbative approach may be applicable to various
physical systems such as the liquid state NMR with a small number of spins
\cite{Nielsen1998,Zhang2005,Alvarez2010}, solid-state NMR
\cite{Cappellaro2007,Fiori2009}, optical lattices \cite{Simon2011}, or quantum
dots \cite{Petta2005}.

\section*{ACKNOWLEDGMENTS}

This material is based upon work supported by NSFC (Grant Nos. 11005099,
11075013, 11175110), Fundamental Research Funds for the Central Universities
(Grant No. 201013037), an Ikerbasque Foundation Startup, the Basque Government
(grant IT472-10) the Spanish MICINN (Project No. FIS2009-12773-C02-02), and the
NSF PHY-0925174, AF/AFOSR No.~FA9550-12-1-0001.

\end{document}